\documentclass[reprint,amsmath,amssymb,aps,prl,groupedaddress,nofootinbib,twocolumn,superscriptaddress]{revtex4-1}
\usepackage{graphicx}
\usepackage{amsthm,amssymb,amsmath,braket,mathdots}
\usepackage{bm}
\usepackage[pagebackref=false,pdfnewwindow=true]{hyperref} 
\usepackage{epstopdf,psfrag}
\usepackage{relsize,amsbsy}
\usepackage[export]{adjustbox}
\usepackage{soul}
\usepackage{graphicx,xcolor,tikz}
\usepackage{float}
\usepackage{subcaption}

\newcommand{\be}{\begin{equation}}
\newcommand{\ee}{\end{equation}}

\newcommand{\bit}{\begin{enumerate}}
	\newcommand{\eit}{\end{enumerate}}

\definecolor{bananayellow}{rgb}{1.0, 0.88, 0.21}
\definecolor{straw}{rgb}{0.32, 0.28, 0.1}

\begin{document}
	\title{Quantum Many-Body Scars in Optical Lattices}
	\author{Hongzheng Zhao}
	\email{hongzheng.zhao17@imperial.ac.uk}
	\author{Joseph Vovrosh}
	\author{Florian Mintert}
	\affiliation{\small Blackett Laboratory, Imperial College London, London SW7 2AZ, United Kingdom}
	\date{\today} 
	\author{Johannes Knolle  }
	\affiliation{Department of Physics TQM, Technische Universit{\"a}t M{\"u}nchen, James-Franck-Stra{\ss}e 1, D-85748 Garching, Germany}
\affiliation{Munich Center for Quantum Science and Technology (MCQST), 80799 Munich, Germany}
\affiliation{\small Blackett Laboratory, Imperial College London, London SW7 2AZ, United Kingdom}

	\begin{abstract}
The concept of quantum {\it many-body scars} has recently been put forward as a route to describe weak ergodicity breaking and violation of the Eigenstate Thermalization Hypothesis.
We propose a simple setup to generate quantum many-body scars in a doubly modulated Bose-Hubbard system which can be readily implemented in cold atomic gases. The dynamics are shown to be governed by kinetic constraints which appear via density assisted tunneling in a high-frequency expansion. We find the optimal driving parameters for the kinetically constrained hopping which leads to small isolated subspaces of scared eigenstates.  The experimental signatures and the transition to fully thermalizing behavior as a function of driving frequency are analyzed. 
	\end{abstract}
	
	\maketitle

{\it Introduction.--} There have been great efforts in the last decades to further our understanding of quantum thermalization, \textit{i.e.}
how local observables of a closed quantum system converge to a stationary state of thermal equilibrium, even though the system follows unitary dynamics.
It is believed, with strong support from extensive numerical studies~\cite{polkovnikov2011}, that quantum thermalization can be sufficiently described by the Eigenstate Thermalization Hypothesis (ETH)~\cite{rigol2008}. Additional evidence comes from the rapid progress in controllable experiments, for example cold atoms~\cite{UQ,Review,bernien2017} and trapped ions~\cite{Blatt2008,blatt2012}, which have enabled the characterization of different paths to thermalization~\cite{trotzky2012}. 

While ETH can explain thermalization for the vast majority of quantum systems, there are also systems substantially deviating from predictions by ETH,
as observed in quench experiments showing that some systems keep some memory of their initial state~\cite{choi2016exploring}.
Prominent examples include integrable systems~\cite{calabrese2011} and many-body localized (MBL) phases~\cite{nandkishore2015,Abanin2019,Kohlert2019} with an extensive number of conserved quantities.
 Beyond the two extremes of fully thermalizing versus fully many-body localized systems, the basic question has emerged whether ergodicity can be weakly broken by special eigenstates rather than the full spectrum.

Recently, an interesting phenomenon suggesting weak ergodicity breaking in non-integrable systems was observed experimentally on a Rydberg atom platform~\cite{bernien2017}. Unexpected persistent coherent oscillations appeared for a set of special initial states. It turns out that kinetic constraints  in the system play an important role in the appearance of slow equilibration; for example the constrained PXP model of a spin-$\frac{1}{2}$ chain~\cite{turner2018weak,choi2019,khemani2019,Bull2019,Lin2019,lin2019slow} displays similar phenomenology.
A special band of non-thermal eigenstates, dubbed {\it quantum many-body scars}, emerge;
they have support in a small portion of Fock-space while other states obey ETH.
Any initial state with a large overlap with scar states {displays}
non-ergodic behavior 
{that manifests itself through persistent coherent oscillations}.

By now, there exist several
theoretical proposals for quantum systems hosting quantum many-body scars; most of them are spin models~\cite{ho2019,choi2019,pai2019,mukherjee2019,schecter2019weak,turner2018weak,michailidis2019,sugiura2019,haldar2019,Pablo2019ergodicity,Sanjay2018,Sanjay2018spinS,Thomas2019,Thomas2020}, but also fermionic~\cite{sanjay2019,Frank2019} and bosonic models~\cite{Ana2019} can show this effect.
Besides the spin-model realized with Rydberg atoms, all of these models are designed such that they feature quantum many-body scars with little consideration on actual implementability.
A crucial question is whether quantum many-body scars can also occur in experimentally feasible settings beyond spin systems.

We propose an experiment consisting of bosonic atoms trapped in an optical lattice,
with temporal modulation of on-site energy and interaction.
In the fast driving regime, this can be described by an effective model~\cite{Flo2,SIT} with density-assisted tunneling processes,
such that the rate for a particle to tunnel depends on the number of particles on its site and neighboring sites.
Since these rates can be tuned in terms of driving amplitudes and frequencies,
optimally chosen driving parameters realize {\it kinetic constraints}, {\it i.e.} selection rules between states that would otherwise be connected via tunneling processes.
For the effective model, the Hilbert space is then separated into disjoint subspaces, leading to the appearance of quantum many-body scars~\cite{Frank2019}.
We show that scarred states are experimentally detectable from suitable initial states that result in coherent oscillations that exist for time-scales exceeding the natural time-scales for thermalization by far. 

{\it Doubly Modulated Model.--} We focus on a system of spinless bosons occupying the lowest band in an {1D} optical lattice described by the Bose Hubbard Model (BHM),
\begin{equation} 
\label{eq.H1}
\hat{H}(t) = \sum_{\langle pq\rangle}\hat{c}_p^{\dagger}J_{pq}\hat{c}_q+\frac{U(t)}{2}\sum_p \hat{n}_p(\hat{n}_p-1)+F(t)\sum_{p}\epsilon_p\hat{n}_p \ , 
\end{equation}
where $\hat{c}_p(\hat{c}_p^{\dagger})$ annihilates (creates) a boson at site $p$, and $\hat{n}_p=\hat{c}_p^{\dagger}\hat{c}_p$ is the occupation number operator. $U(t)$ is the amplitude of the on-site interaction, $\epsilon_p=-p$ represents the tilt potential, $F(t)$ is the shaking amplitude,
and {$J_{pq}=-J$} is the bare tunneling rate between nearest neighbors. 
We will consider the doubly modulated version of the BHM~\cite{EU,Zhao2019}, in which both $U(t)$ and $F(t)$ are modulated periodically with period $T$.
In practice, the time dependence results from a periodic tilt or acceleration of the lattice and 
 modulation of a magnetic field inducing a Feshbach resonance~\cite{FR4,FR3}.
We exploit the freedom to choose suitable driving profiles to realize kinetically constrained hopping of bosons on a chain.
It turns out that a monochromatically modulated on-site interaction
$
U(t) = U_d\cos(\omega t)
$
and a shaking in the form
$
F(t)=F_2\cos(2\omega t)+F_4\cos(4\omega t)
$
are all that is required.
To simplify notation, we will also use the dimensionless quantities $\tilde U_d=U_d/\omega$, $\tilde F_2=F_2/2\omega$ and $\tilde F_4=F_4/4\omega$ in the following.

\textit{Kinetically Constrained Hopping.--} 
{The Hamiltonian $\tilde{H}(t)$ in the frame induced by the driving term of the Hamiltonian admits a perturbative high-frequency expansion even for strong driving~\cite{Zhao2019, EU}.
The dynamics can then be approximated by a time-independent effective Hamiltonian $\hat{H}_{\mathrm{eff}}$~\cite{HF}.}
The lowest order contribution to the high-frequency expansion reads
$\hat{H}^0_{\mathrm{eff}} = \frac{1}{T}\int_{0}^{T}dt\tilde{H}(t)$,
which in the present case reduces to the so-called density assisted tunneling 
\be
\hat{H}^0_{\mathrm{eff}} = \sum_{pq}\hat{c}_p^\dagger\hat{A}^0_{pq}(\hat{n}_p,\hat{n}_q)\hat{c}_q\ , 
\label{eq.full_Hamiltonian}
\ee
where the density dependence is  contained in the operator
\begin{equation}
\hat{A}^0_{pq}(\hat{n}_p,\hat{n}_q)=-J
 \mathcal{J}_0\Big(\tilde U_d(\hat{n}_p-\hat{n}_q),\tilde F_2(p-q),\tilde F_4(p-q)\Big)\ ,
\label{eq.A}
\end{equation}
in terms of the {zeroth order} three-dimensional Bessel function {$\mathcal{J}_0$}~\cite{VerdenyBessel}.
The existence of the term $\hat n_p-\hat n_q$ in the operator $\hat{A}^0_{pq}$ indicates that the tunneling rates depend on the occupation number difference of specific initial and final states~\cite{EU,Zhao2019}.	
For example, in a chain with the initial Fock state $\ket{\psi_i}=\ket{n_1\hdots n_N}$ one obtains
\be
\hat{c}_p^{\dagger}\hat{A}^0_{p,p+1}\hat{c}_{p+1}\ket{\psi_i}=h_{n_p,n_{p+1}}^L\hat{c}_p^{\dagger}\hat{c}_{p+1}\ket{\psi_i}\ ,
\ee
with the prefactor
\begin{equation}
h^L_{n_p,n_{p+1}}=- J\mathcal{J}_0(\tilde U_d(-n_p+n_{p+1}-1),\tilde F_2,\tilde F_4)\ ,
\end{equation} 
that can be interpreted as the tunneling rate for a particle to tunnel from site $p+1$ to $p$, {\it i.e.} to the left.
Similarly the  rate for a particle to tunnel from site $p$ to $p+1$ reads
\begin{equation}
h^R_{n_p,n_{p+1}}=- J\mathcal{J}_0(\tilde U_d(-n_p+n_{p+1}+1),\tilde F_2,\tilde F_4)\ .
\end{equation}
As states with low occupations per site will be particularly relevant,
the following rates
\begin{eqnarray}
\label{equ.channel}
\begin{aligned}
h^L_{(0,1)}&=h^R_{(1,0)}=h_{(2,1)}^R= h_{(1,2)}^L =-J\mathcal{J}_0(0,\tilde F_2,\tilde F_4)\ ,\\
h_{(1,1)}^{L/R} &=h_{(2,0)}^{R}=h_{(0,2)}^{L}=- J\mathcal{J}_0(\tilde U_d,\tilde F_2,\tilde F_4)\ ,\\
h_{(2,1)}^L &= h_{(1,2)}^R=h_{(0,3)}^L= h_{(3,0)}^R=- J\mathcal{J}_0(2\tilde U_d,\tilde F_2,\tilde F_4)\ ,\\
\end{aligned}
\end{eqnarray}
play an important role in the formation of many-body scars.

\begin{figure}[t]
		\centering
			\begin{subfigure}{0.49\textwidth}
			\centering
			\includegraphics[width=0.59\linewidth]{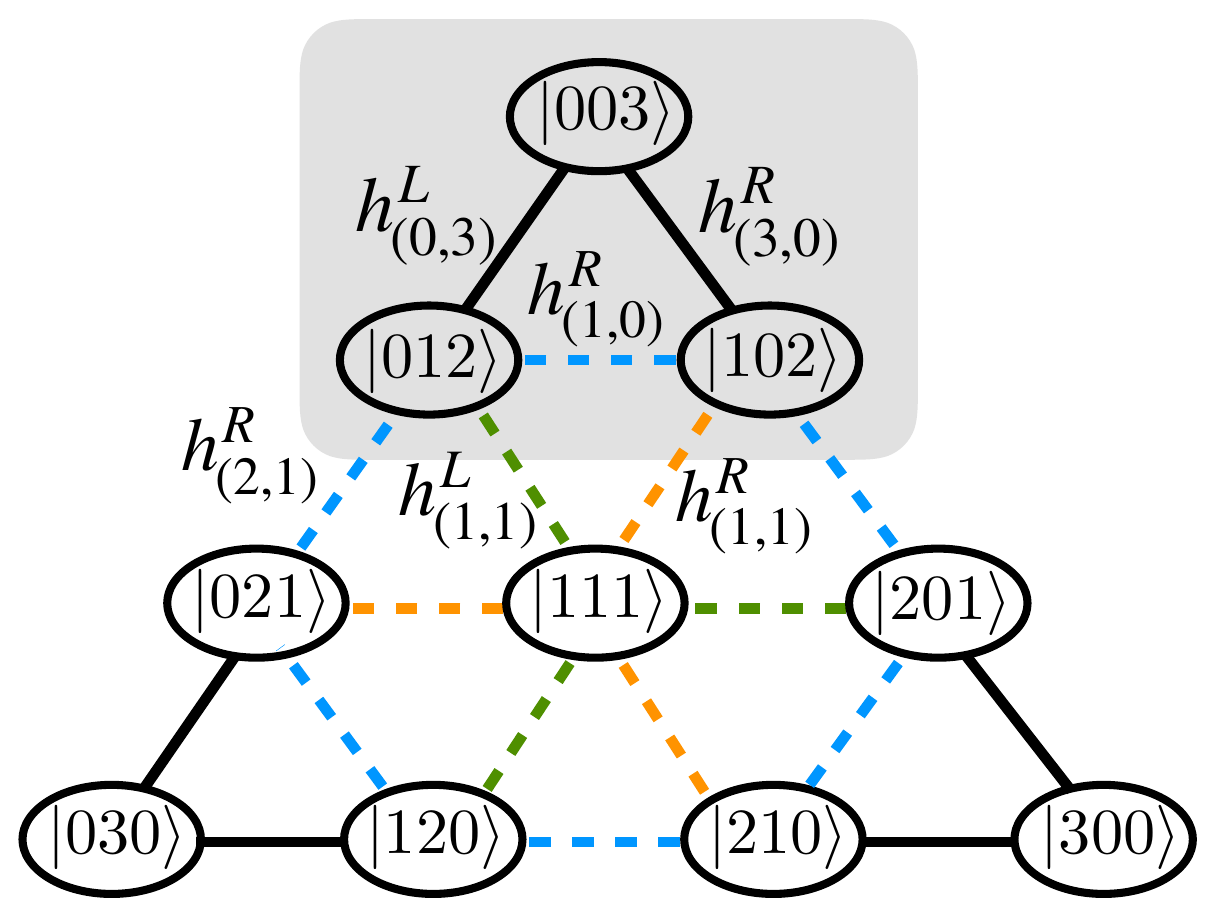}
		\end{subfigure}
	\caption{Fragments for a system of 3 sites. Connections are assigned with hopping rates. If all dashed ones are forbidden for conditions in Eq.~\eqref{eq.condition}, four blocks of states appear. For clarity one of these is shadowed in gray. }
		\label{fig:fig3sites}
\end{figure}

{\it Fragmentation in Fock-Space.--}
In order to understand how specific choices of the rates in Eq.~\eqref{equ.channel} result in the desired fragmentation of Hilbert space, it is instructive to consider the basic example of three lattice sites with periodic boundary condition as illustrated in Fig.~\ref{fig:fig3sites}.
Tunneling processes between different Fock states are indicated by solid and dashed lines,
and processes that occur with the same rate are depicted with the same color.

If all the rates of processes depicted with dashed lines vanish, then the Hilbert space is being fragmented into four blocks -- one block containing the states $\ket{012}$, $\ket{102}$, $\ket{003}$, two blocks containing their cyclic permutations, and the one-dimensional block $\ket{111}$. 
This fragmentation naturally occurs if the rates specified in Eq.~\eqref{equ.channel} satisfy the conditions
 \begin{eqnarray}
 \begin{aligned}
 h_{(2,0)}^{R}=h_{(0,2)}^{L}=h^{L/R}_{(1,1)}=0\ ,\  h^L_{(1,2)}=h^R_{(2,1)}=0\ .
 \label{eq.condition}
 \end{aligned}
 \end{eqnarray}
In practice it is not necessary that these rates vanish exactly, but clear deviation from ETH can also be observed in the presence of {\it leaking channels}, {\it i.e.} small, but finite amplitudes of these rates.

Fragmentation in systems of any number of sites and filling factor can be understood in terms of elementary building blocks of few sites. One exemplary building block contains only the Fock states $\ket{n_1n_2n_3n_4n_5n_6}$ for a system of six atoms on six sites, with $n_1=n_6=0$, $n_2\le 2$ and $n_5\le 2$, for instance the state $\ket{\psi} = \ket{003300}$. Due to the constraints in Eq.~\eqref{eq.condition}, the vanishing occupation on sites 1 and 6 in $\ket{\psi}$ is conserved throughout time evolution. In turn, for a products of such 6-site states, there will be no mixing in-between thus they form a fragment which is much smaller than the total Hilbert space. In practice, there are many more elementary building blocks, giving rise to a substantial number of scarred states in large systems.

\begin{figure}[t]
	\centering
	\begin{subfigure}{0.49\textwidth}
		\centering
		\includegraphics[width=0.99\linewidth]{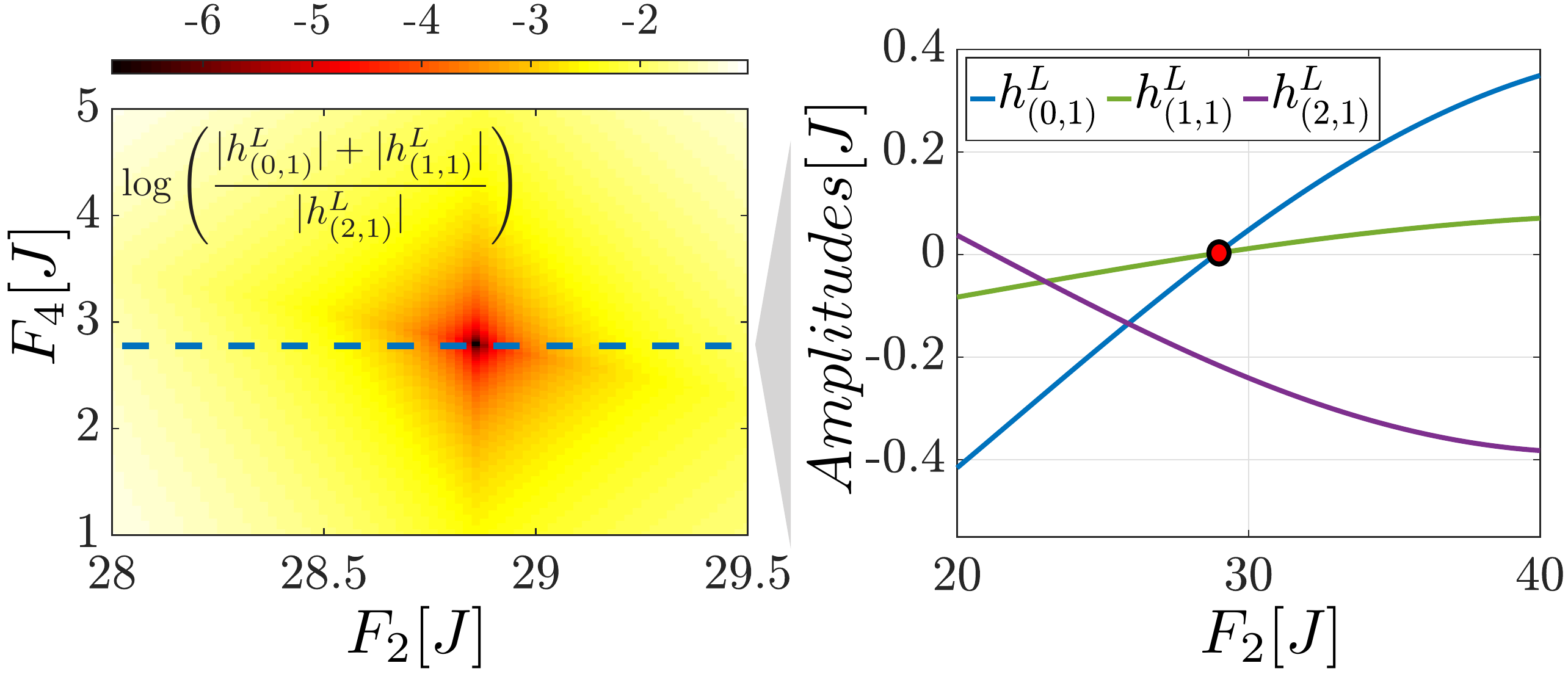}
	\end{subfigure}
	\caption{Left: Figure of merit for the appearance of quantum many-body scars given by the ratio (log scale) of leaking and hopping channels for a Hilbert space fragment. The red region labels the best parameter space with ratios lower than $10^{-4}$. Right: Hopping rate as a function of $F_2$. Two leaking channels cross zero at the same time.}
	\label{fig:amplitude}
\end{figure}
{\it Optimal Control of Scars.--} 
In order to realize the above scars experimentally, the driving parameters need to be chosen such that
 \begin{eqnarray}
 \mathcal{J}_0(\tilde U_d,\tilde F_2,\tilde F_4)=0\ ,\ \   \mathcal{J}_0(0,\tilde F_2,\tilde F_4) = 0\ ,
 \label{eq.conditionParameters}
 \end{eqnarray}
 to satisfy the conditions given in Eq.~\eqref{eq.condition}.
 In general, one needs a large modulating interaction $U_d$, to realize a notable difference between hopping channels (Eq.~\eqref{equ.channel}) with different particle numbers. For concreteness, we fix $U_d/J=12, \omega/J=6$  and show the ratio between dominant leaking and leading hopping rates within the fragment  $(|h_{(0,1)}^L|+|h_{1,1}^L|)/|h_{(2,1)}^L|$ as a function of $F_2,F_4$ in the left panel of Fig.~\ref{fig:amplitude} . This is a suitable figure of merit for the appearance of long time coherent oscillations as the inverse of the ratio captures the time scale for the leaking effects to become notable. There is a sweet spot in black around $(28.8,2.8)$ with ratios below $10^{-6}$. In the right panel, the rates are depicted with fixed {$F_4 =2.8J$}, and one can see two leaking rates crossing zero at the same point. Therefore those two channels can be tuned to both be strictly forbidden, leading to perfect fragments. 

So far we have restricted our discussion on the conditions for fragmentation of the Hilbert space of the effective Hamiltonian $\hat{H}^0_{\mathrm{eff}}$, which drastically changes the  stroboscopic dynamics of the driven Bose-Hubbard chain in the high-frequency regime, but higher order processes of magnitude {$O(1/\omega)$} will play an increasing role for smaller frequencies. This will inevitably open leaking channels and a transition to a thermalizing system without scars is expected. 
In the following, higher order effects are analyzed via the spectrum of the full Floquet operator, $\hat{U}(T) = \mathcal{T}\int_{0}^{T}\exp \left(-i \hat{H}(t) dt\right)$, obtained via exact diagonalization for unit-filling. The ratio between the driving amplitude $U_d,F_2,F_4$ and frequency $\omega$ is fixed to the optimal point discussed above.

\begin{figure}[t]
	\centering
	\begin{subfigure}[b]{.49\linewidth}
		\centering
		\includegraphics[width=0.99\textwidth]{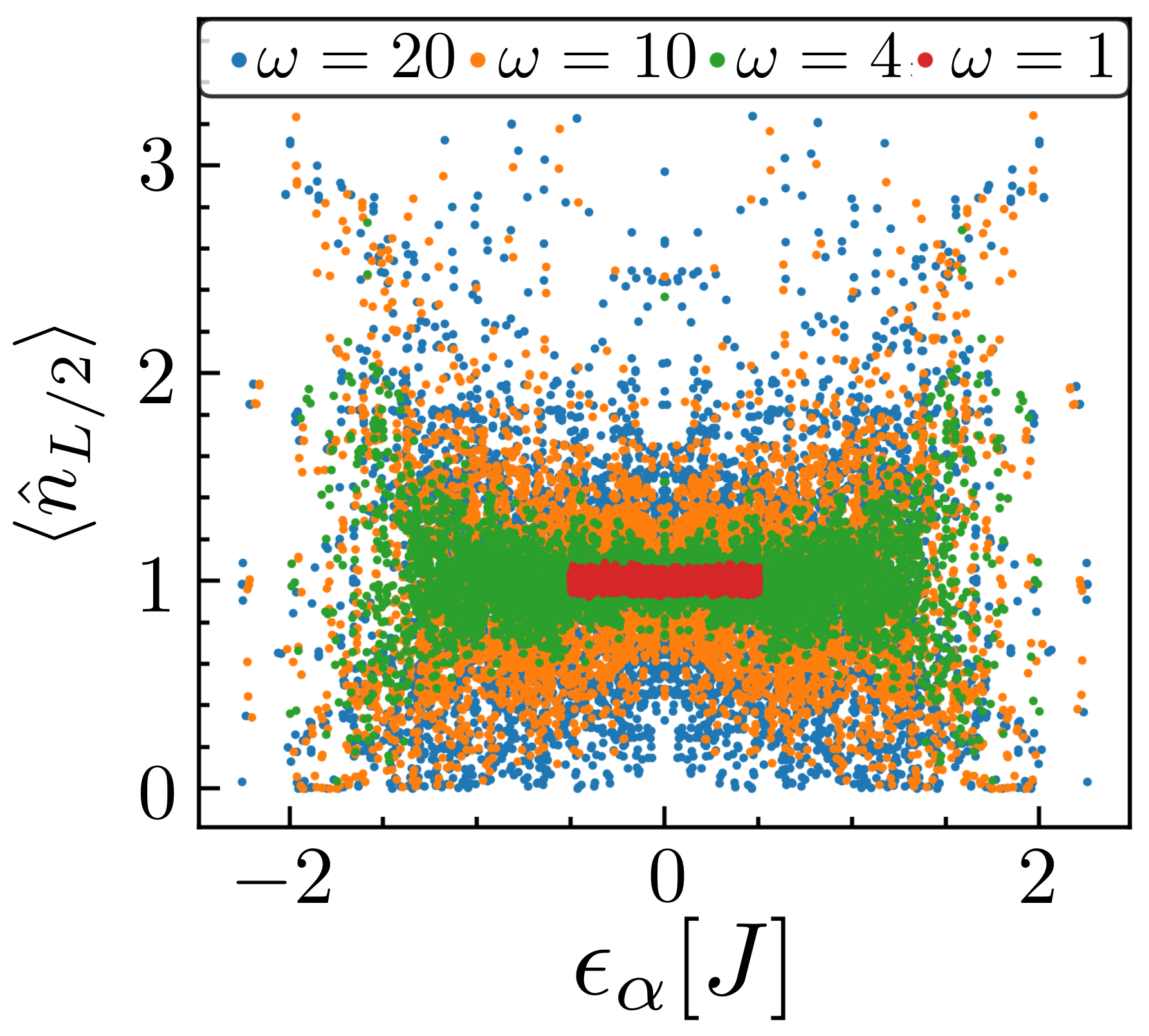}
		\caption{}
		\label{fig:entObs}
	\end{subfigure}
	\begin{subfigure}[b]{.48\linewidth}
		\centering
		\includegraphics[width=0.99\textwidth]{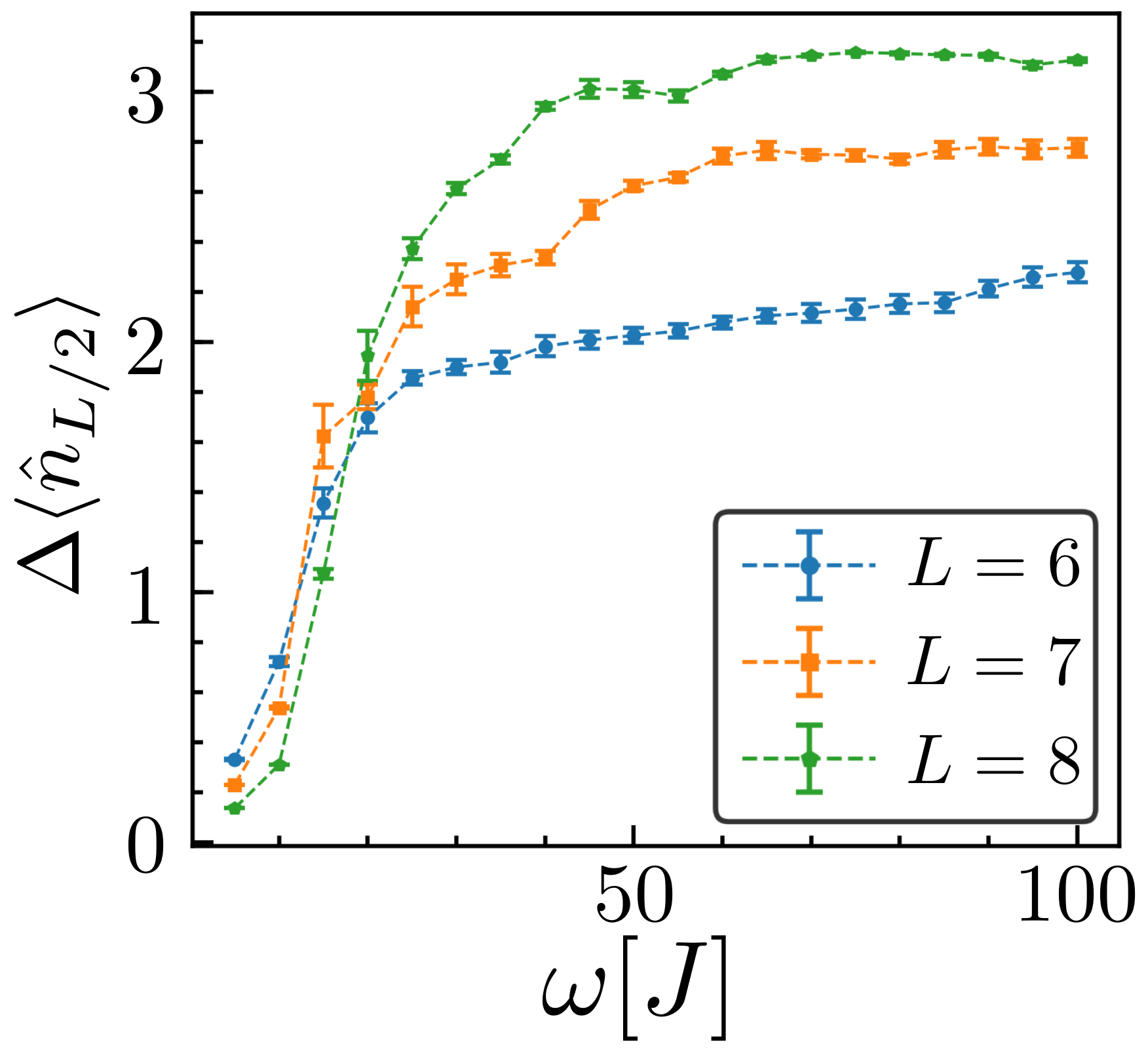}
		\caption{}
		\label{fig:entscale}
	\end{subfigure}

	\caption{Violation of ETH as a function of driving frequency {in units of $J$}. (a) Expectation value of the local particle number operator, with respect to eigenstates of the Floquet operator as function of the quasienergy for system size $L=8$.	
	 (b) Width of the distribution of $\langle \hat{n}_{L/2} \rangle$ as a function of frequency for different system sizes .}
\end{figure}

In contrast to
the smooth dependence of expectation values of local observables as a function of eigenstates' energy,
scarred states typically lead to very different values even for states close in energy~\cite{Pablo2019ergodicity}.
This violation of ETH can be seen in Fig.~\ref{fig:entObs} depicting the expectation value of the local particle number $\langle \hat{n}_{L/2}\rangle$ for eigenstates of the Floquet operator as a function of their quasienergy.
In the case of slow driving, depicted in red, the local particle number is close to the value of one, with essentially no dependence on the quasienergy $\epsilon_\alpha$.
For faster driving, however, the local particle number changes seemingly chaotically from one eigenstate to the next.
The transition to the fast driving regime can also be appreciated from the range of quasi energies: outside the fast driving regime ($\omega/J=1$), the full interval $[-\omega/2,\omega/2)$ is occupied, whereas the occupied interval hardly grows with increasing $\omega$ for $\omega/J\ge 4$.

The width of the distribution of local particle numbers is best characterized in terms of the difference between the typical largest and typical smallest local particle number.
Fig.~\ref{fig:entscale} depicts
the width $\Delta \langle \hat{n}_{L/2}\rangle=\langle \hat{n}_{L/2}\rangle_{\mathrm{max}}-\langle \hat{n}_{L/2}\rangle_{\mathrm{min}}$,
where $\langle \hat{n}_{L/2}\rangle_{\mathrm{max/min}}$ is the average over the $20$ largest/lowest local particle numbers.
The error bars are estimated from the variances of these averages.
The dependence of $\Delta \langle \hat{n}_{L/2}\rangle$ on the driving frequency $\omega$, shows a well-pronounced increase indicating the formation of many-body scars on the transition to the high-frequency regime.

Having established the appearance of quantum many-body scars, the last -- but crucial -- aspects to discuss are their experimental signatures and realization.

{\it Coherent Oscillations.--}  A direct way of detecting scarred states is by studying the long time oscillations of specially prepared initial states. 
For the optimal parameters in the high-frequency limit, the state will remain in the subspace of its fragment and thus show coherent oscillations.
If the driving frequency is finite, but sufficiently large so that the rates of leaking channels are much smaller than tunnelings rates within the subspace, these coherent oscillations will be damped out slowly and their lifetime can exceed the thermalization time by far.

\begin{figure}[t]
	\centering
	\begin{subfigure}[b]{0.49\linewidth}
		\centering
		\includegraphics[width=0.97\textwidth]{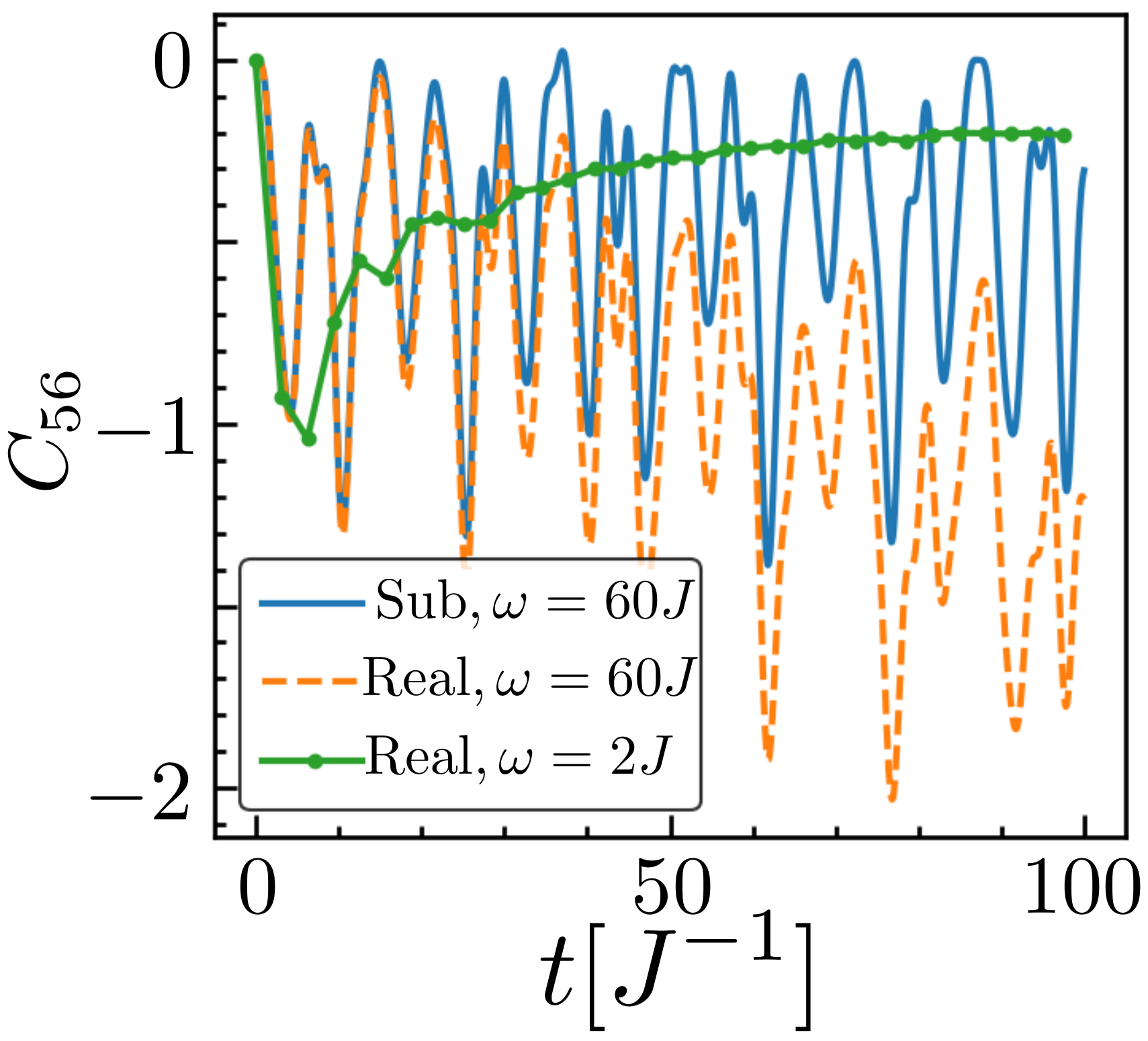}
		\caption{}
		\label{fig:correlation}
	\end{subfigure}
	\begin{subfigure}[b]{0.49\linewidth}
		\centering
		\includegraphics[width=0.99\textwidth]{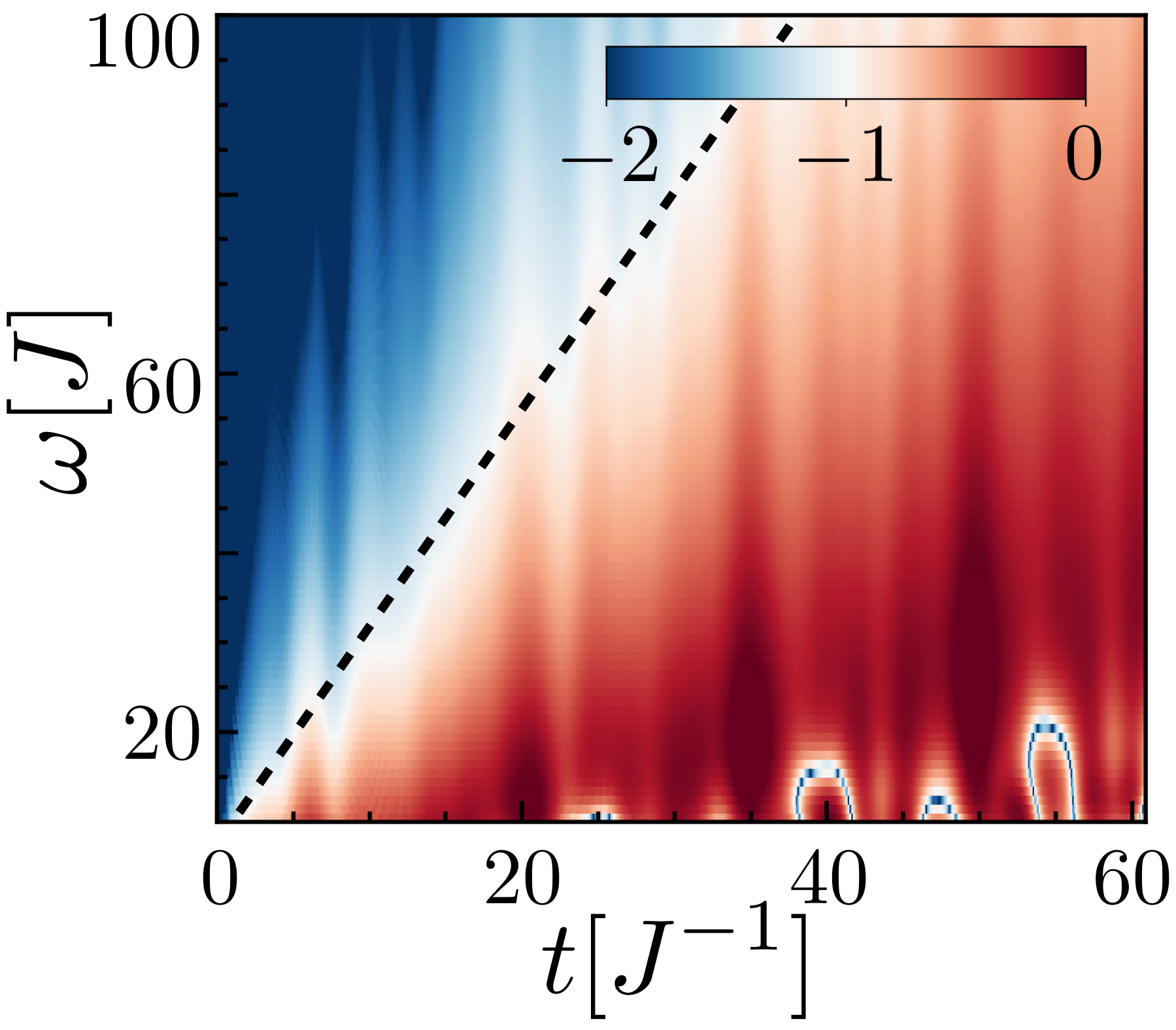}
		\caption{}
		\label{fig:correlation2D}
	\end{subfigure}
	\caption{(a) Correlation function given optimal parameters. Coherent oscillation with fast driving matches well with subsystem approximation, indicating the existence of scared states. In contrast, correlation saturates quickly in the ergodic case with $\omega=2J$ where higher order effects induce deleterious leaking. (b) Correlation difference $\log_{10}(R(t))$ between real dynamics and subspace approximation (Eq.~\eqref{eq:ratio}) for varying frequencies. Some blue spots appear within the red region at longer times because of the accidental similarity between $C_{\mathrm{real}}(t)$ and $C_{\mathrm{sub}}(t)$.}
\end{figure}

As a specific example,
we focus on the stroboscopic dynamics of a system with ten sites with periodic boundary conditions and initialized in the Fock state $\ket{1100330011}$.
The coherent oscillation
of the two-point correlation function
${C_{56}}(t) =\langle \hat{n}_5(t)\hat{n}_6(t)\rangle-\langle \hat{n}_5(t)\rangle \langle \hat{n}_6(t)\rangle$,
of the two central sites {is} depicted in Fig.~\ref{fig:correlation}.
The actual system dynamics $C_{\mathrm{real}}(t)$ are depicted in orange and compared to the dynamics $C_{\mathrm{sub}}(t)$
induced by the subsystem Hamiltonian~\cite{Turner2018} in which all rates of leaking channels are set to vanish. 
The ergodic case with slow driving is depicted in green, and it shows that the correlation saturates rapidly on a time-scale $t\sim O(1/J)$.
While the comparison between actual scar dynamics and the effective subsystem dynamics gives evidence of damping due to leakage,
the oscillations in the scarred dynamics also clearly exceed the time-scale of thermalization.

In order to turn this qualitative observation into a quantitative assessment of the lifetime of the coherent oscillations, it is helpful to define
the difference between the two correlation functions
\begin{eqnarray}
\label{eq:ratio}
R(t) = {|{C_{\mathrm{real}}(t)-C_{\mathrm{sub}}(t)}|}, 
\end{eqnarray}
comparing the actual dynamics and the effective subsystem dynamics. 
The difference $R(t)$ is depicted in $\log10$ scale in Fig.~\ref{fig:correlation2D} as a function of time $t$ and driving frequency $\omega$.
As one can see, the difference is very small (blue) for short times, and there is a clear increase (red) for longer times.
The point in time around which this transition occurs grows linearly with $\omega$, as indicated by a dashed black line.
This allows us to define lifetimes 
for scar states.
The proportionality to $\omega$ indicates that the scar states become increasingly stable as the fast driving regime is approached ($\omega\to\infty$).

 Similar long-lived coherent dynamics,  {\it e.g.} Bloch oscillations (BO), also appear in a static tilted lattice without driving~\cite{Geiger2018,buchleitner2003}. However, there are several crucial differences between BO and scar dynamics that can be employed to distinguish them, {\it e.g.} initial state dependency and stability to additional interactions, see more details in the Supplementary Material.

{\it Experimental Realization.--} The doubly modulated Hubbard system in cold atomic gases and the required initial states can be constructed via existing experimental techniques. The driving can be achieved by periodically modulating external magnetic fields in the vicinity of Feshbach resonances~\cite{FR4,ExpCorrelatedHopping} and lattice shaking~\cite{PDMBL}.
In practice, there will be an additional, typically harmonic trapping potential; since it commutes with the driving term in the Hamiltonian (Eq.\eqref{eq.H1}), it does not contribute to the kinetic constraints resulting in the scar states.
The remaining challenge is {thus} to prepare fine tuned initial states. The easiest states to verify fragmentation and violation of ETH are the Mott insulator state and the density wave state with two particles on every second site. Both of them are eigenstates of the effective Hamiltonian, thus, they will be frozen in the presence of fast driving. For more interesting cases discussed above, one can first prepare Mott states with three bosons per site~\cite{Mott} and 
 then apply the single-spin addressing scheme~\cite{single-spin-addressing} to remove unwanted particles on selected sites to obtain for example $\ket{\dots003300003300\dots}$. This state provides an example which shows coherent oscillations for an exceptionally long time before thermalization in the high-frequency limit.

{\it Conclusions \& Outlook.--} 
Our concrete proposal enables the experimental realization of quantum many-body scars in a bosonic system, allowing the study of thermalization and lack thereof in systems substantially larger than those possible to study in any theoretical analysis. The Floquet engineering of kinetically constrained hopping discussed here can be achieved independently of system dimensionality and geometry, thus paving the way to substantially deepen our understanding of different thermalization paths in two- and three-dimensional systems with different lattice geometries (some of which can result in topological order~\cite{ok2019topological}).

The suppression of leaking channels, as well as kinetically constrained hopping in general, is not only helpful for enabling experiments for quantum many-body scars, but it can potentially also be used to suppress processes like heating~\cite{AchilleasHeating} that poses a serious limitation to quantum simulations based on driven atomic gases.
Finally, the ability to open and close selected tunneling channels also opens up new pathways for state preparation like the accurate creation of a coherent superposition of different Fock states~\cite{StateEngineering} as required {\it e.g.} for precision sensing.

{\it Acknowledgements.--}
We acknowledge helpful discussions with Monika Aidelsburger, Pablo Sala and Ana Hudomal.

\bibliography{citations_Scars}
\bibliographystyle{unsrt}

\newpage
\appendix
\section{Comparison between Bloch Oscillations  and Scar Dynamics}
With tilted lattices, non-interacting atoms can show long-lived Bloch oscillation (BO)~\cite{Geiger2018,buchleitner2003}. However, there are several crucial differences between BO and the coherent scar dynamics(Fig.~\ref{fig:correlation}) which can be employed to distinguish them experimentally. 

First, in contrast to BO, the existence of the coherent oscillation presented in the main content strongly depends on the specially selected initial conditions. Only properly prepared initial states,  {\it e.g.} Fock states constructed with building blocks, can exhibit coherent dynamics. Alternatively, the Mott and density wave states are dynamically frozen for fast driving. As a counter-example, the superfluid state will thermalize very quickly. Moreover, the oscillating frequency of BO is fixed by the tilted potential \cite{Geiger2018}, while for the coherent scar dynamics, it can vary by choosing different initial states.

Second,  in the presence of additional interactions (of energy scale $U$ comparable to bare tunneling $J$), BO are very sensitive to interaction-induced dephasing and instabilities thus decay rapidly~\cite{buchleitner2003}. Yet this is not true for scar dynamics. Interactions commute with the unitary transformation to the rotating frame thus can be simply added to the lowest order effective Hamiltonian (Eq.~\ref{eq.full_Hamiltonian}) as it is. Therefore, the density assisted tunneling will not be impacted and the kinetic constraints leading to Hilbert space fragmentation remain unchanged. Considering higher-order processes, our previous work~\cite{Zhao2019} showed that only negligible $(1/\omega^2)$ effects are induced, therefore the coherent oscillation will still persist for a long time with fast driving.

Third, the correlation between different building blocks serves as a good measure since no tunnelling is permitted in-between in the effective Hamiltonian. As an example, for the initial state $\ket{1100330011}$, the correlation function ${C_{15}}(t) =\langle \hat{n}_1(t)\hat{n}_5(t)\rangle-\langle \hat{n}_1(t)\rangle \langle \hat{n}_5(t)\rangle$ remains zero for a sufficiently long time with fast driving. On the contrary, BO will build up such a spatial correlation after a short period.

All these three aspects exploit the many-body character of the density assisted tunneling which does not have a counterpart in non-interacting systems. Importantly, they are experimentally verifiable via local occupation number or two-point correlation functions. 

\end{document}